\documentstyle[12pt,aasms]{article}
\journalid{337}{15 January 1989}
\articleid{11}{14}
\begin{document}

\title{TRANSIENT GAMMA-RAY SPECTROMETER OBSERVATIONS OF GAMMA-RAY 
LINES FROM NOVAE.  III. THE 478 keV LINE FROM $^{7}$Be DECAY}

\author{M. J. HARRIS\footnote{Universities Space Research Association,
harris@tgrs2.gsfc.nasa.gov}, B. J. TEEGARDEN, 
G. WEIDENSPOINTNER\altaffilmark{1}, 
D. M. PALMER\altaffilmark{1,}\footnote{Present address: NIS-2, D346,
Los Alamos National Laboratory, Los Alamos, NM 87544}, 
T.~L. CLINE, N. GEHRELS, AND R. RAMATY\footnote{Deceased}}
\affil{Code 661, NASA/Goddard Space Flight Center, Greenbelt, MD 20771}

\begin{abstract}

The {\em Wind\/} spacecraft carrying the Transient Gamma Ray Spectrometer 
(TGRS) moves in an extremely elliptical orbit that largely avoids Earth's
trapped radiation belts and albedo $\gamma$-radiation.  The TGRS
therefore enjoys a relatively low level of background that is also
extremely stable.  We show how this stability enables modeling of the
time variability of background lines, which in turn enables a novel 
technique of background subtraction to be used in the detection of transient
astrophysical lines.  We apply a simple version of this method to
the line at 478 keV that is
expected to arise from nucleosynthesis of $^{7}$Be in nearby 
novae.  This search covers the entire southern ecliptic hemisphere
during 1995--1997, including five known individual events, and possible
undiscovered individual events.
The TGRS design also uses {\em Wind}'s 3~s rotation period to modulate
signals from the Galactic center (GC).  We use this feature of the instrument
to search for a quasi-constant level of 478 keV emission from the
accumulation of $^{7}$Be from several novae that are expected to occur
in the direction of the GC during that isotope's 53 d
half-life.  We derive upper limits on the transient (single-nova)
emission which improve on previous limits by about an order of magnitude,
and limits on the steady (many-nova) emission which represent a factor
2 improvement.  Only weak limits can be placed on the key parameters in
the nucleosynthesis and ejection of $^{7}$Be, however.

\end{abstract}

\clearpage

\section{Introduction}

Although clasical novae occur rather frequently in our Galaxy (some tens
of events per year), they are not
expected to contribute greatly to the chemical enrichment of the Galaxy
because they eject relatively small masses of nuclear-processed material.
The rare CNO nuclei $^{13}$C, $^{15}$N and $^{17}$O are exceptions, and so
is the light isotope $^{7}$Li, which is the topic of the present study
(Hernanz \& Jos\'{e} 2000).  About 10\% of the present Galactic abundance
of $^{7}$Li may have come from novae, which are one of the several sources
(mainly cosmic ray spallation and neutrino spallation in 
supernovae: Cass\'{e}, Vangioni-Flam \& Audouze 2001) that raise the
$^{7}$Li abundance above the cosmologically-interesting value left by 
Big Bang nucleosynthesis.  Nova nucleosynthesis of $^{7}$Li is also of
interest in that the $^{7}$Li is produced as unstable $^{7}$Be by the
reaction $^{3}$He($\alpha$,$\gamma$)$^{7}$Be, where $^{3}$He is another
Big Bang isotope whose subsequent evolution must be accounted for.  The
$\beta$-decay of $^{7}$Be with half-life 53.28 d
is accompanied in 10.5\% of cases by emission 
of a $\gamma$-ray of energy 478 keV that potentially provides a direct 
measurement of the nova contribution if it can be detected.

Monitoring of the $\gamma$-ray sky for novae is important because many
events are never detected ($\sim 3$ out of several dozen per year are
discovered, often by amateur astronomers: Harris et al. 1999, 2000, 
hereafter Papers I and II), raising the possibility that original 
discoveries of optically-undetected novae may be made by their 
$\gamma$-ray emission alone.  This requires
experiments with broad fields of view carried on space platforms over
periods of the order of years.  Data from four such experiments have been
analyzed with this purpose in view: the Gamma Ray Spectrometer on the
{\em Solar Maximum Mission\/} ({\em SMM\/}, 1980--1989: see e.g. 
Leising et al. 1988);
BATSE and COMPTEL on the {\em Compton\/} Observatory (1991--2000: Hernanz 
et al. 2000 and Iyudin et al. 1995, 2001); and the Transient 
Gamma Ray Spectrometer (TGRS) on board
{\em Wind\/}, from which we will present results from 1995--1997 in this
paper.  The properties of the TGRS detector, which we describe in \S 3,
are very suitable for monitoring line emission from novae.  The detector
has very good energy resolution, which enables Doppler-shifted lines from
cosmic sources to be distinguished from background lines at the rest
energy; we previously exploited this property in searches for the 511 keV
positron annihilation line, which is blueshifted by a few keV in
novae (Paper I, Paper II).  In this paper we exploit another
property of TGRS to search for the 478 keV line, namely the remarkable
stability of the instrument's background $\gamma$-ray line spectrum
(\S 4).

\section{Properties of Classical Novae}

Our analysis relies to some extent on the expected properties of the
$\gamma$-ray line from the nova event.  The most important factor in
the nucleosynthesis is the composition of the white dwarf upon which
the explosion occurs, since material is expected to have entered the burning
layer by diffusion upward and to have altered its composition substantially
from what was accreted.  Most common are expected to be degenerate CO 
remnants of $\le 1.25~M_{\sun}$ left by lower mass red giants.  More massive 
objects up to the limiting stable mass $1.4~M_{\sun}$ probably have ONe
composition.  It is not clear where the transition in mass between CO
and ONe occurs, or whether indeed there is a range of overlap.  Although
the general white dwarf mass distribution peaks around $0.6~M_{\sun}$,
the distribution of white dwarf masses in novae is skewed towards
higher masses, because the frequency with which nova 
outbursts recur is a steeply rising function of mass.
The ONe remnants and the more massive CO objects $> 1~M_{\sun}$ 
are therefore over-represented.  A ratio of 2:1 for the two subclasses 
CO:ONe is often assumed.

Abundance analyses are not available for most novae.  Properties
may be deduced (with great uncertainty) from the correlations of white
dwarf subclass with other observables.  It appears that explosions on
CO white dwarfs are less energetic.  The ejected masses are of order
$10^{-4}~M_{\sun}$ for both subclasses, but CO novae have lower expansion
velocities $\sim 1000$ km s$^{-1}$ (Warner 1995).  One of the individual
novae that we will consider, the ONe type CP Cru, had a velocity of
2000 km s$^{-1}$ (Della Valle 1996).

Theoretical models predict that CO novae are the strongest sources of
the 478 keV $\gamma$-ray line (Gom\'{e}z-Gomar et al. 1998, Hernanz \&
Jos\'{e} 2000) since they may produce about $10^{-10}~M_{\sun}$ 
of $^{7}$Be per event.  However, predictions of the $^{7}$Be
mass fraction produced vary considerably (see e.g. Starrfield et al. 
2000a, Hix et al. 2000, and Hernanz \& Jos\'{e} 2000).
Further uncertainty is added by the failure (in
general) of models to reproduce the full $10^{-4}~M_{\sun}$
of ejecta, falling short by an order of magnitude or more.  The
ejecta velocities are also usually underestimated.  

Distance estimates for novae are highly uncertain.  
An empirical relation exists between rate of decline and absolute
visual magnitude:
\begin{eqnarray}
M_{V} & = & 2.41 \log t_{2} -10.7 ~~~\mbox{for 5 d $< t_{2} < 50$ d} \\
 & = & -9 ~~~\mbox{for $t_{2} \le 5$ d} \\ 
 & = & -6.6 ~~~ \mbox {for $t_{2} > 50$ d}
\end{eqnarray}
(Warner 1995), where $t_{2}$, the speed
class, is the time taken for $m_{V}$ to increase by 2 from
discovery.\footnote{ A fuller description of our assumptions and
input values to this formula, plus references, is found in Paper I.  
In two cases (BY Cir
and V888 Cen) our result is in very good agreement with that of
Shafter (1997) obtained by a variation of the same method.  Issues
such as visual extinction are common to this method and the 
Eddington luminosity method described next.}  

Following the explosion in and ejections by thermonuclear runaway,
the remaining surface layers undergo hydrostatic H burning on a
longer time-scale $\sim 1$ yr (Shore, Starrfield, \& Sonneborn 1996)
and a convective envelope developes.
During this time the bolometric luminosity is thought to remain
constant at the Eddington value $L_{edd}$ (Warner 1995), but the
peak of the emission moves from the optical to the UV as expansion
reveals hotter underlying layers.  This leads to a second method
of estimating distance that has sometimes been used (e.g. Stickland et 
al. 1981, Evans et al. 1990).  Let it be assumed that at visual maximum
almost all the flux is at visual wavelengths [as Stickland et al.
found for V1668 Cyg (1978)].  The absolute visual magnitude is then
known from the Eddington formula 
$L_{Edd} = 1.3 \times 10^{38} \mu M_{WD}$ erg s$^{-1}$, where $\mu$
is the mean molecular weight and $M_{WD}$ the white dwarf mass in
$M_{\sun}$; the distance follows from a comparison of apparent and 
absolute visual magnitudes.

Post-outburst behavior raises further questions about 
$^{7}$Be observability, which depends on the time-scale on which the
expanding envelope becomes transparent to the $\gamma$-ray line;
it obviously must be well within the 53 d half-life of $^{7}$Be if
the line is to be seen.\footnote{ Models which directly
accelerate a large fraction of the ejecta in the explosion
can fulfil this condition (Gom\'{e}z-Gomar et al. 1998).  
Another process acting is the ejection
of the $^{7}$Be in a low-velocity, late-time wind of which part is
accelerated by a drag force due to the companion when a common
envelope developes between primary and secondary (Livio et al. 1992).  
It was applied to $\gamma$-ray line emission in novae V838 Her
(Starrfield et al. 1992, 1993) and V1974 Cyg (Starrfield et al. 
1993).  Models of this type will not become transparent to
$\gamma$-rays for $\ge 100$ d or two half-lives, severely
depleting the detectable $^{7}$Be.}  We assume that this is the case, 
following Gom\'{e}z-Gomar et al. (1998), who predict $\gamma$-ray 
transparency after 5--13 days, depending on the mass and composition 
of the model.  These models also predict large yields of $^{7}$Be 
and 478 keV line fluxes.  Our measurements can therefore
be regarded as an approach to testing these "optimistic" models.

During 1995--1997 five novae were discovered in the southern
hemisphere.  They are listed in Table 1, with composition
information where available, and distance estimates
from the two empirical methods described above.  There are
obvious sources of uncertainty in the distance measurements,
notably the time of visual maximum (before discovery) and, in the
case of the $L_{Edd}$ method, possible neglect of a bolometric
correction and super-Eddington velocities in some novae (Schwarz 1999).  
In addition, there is a single reliable distance measurement 
(3180 pc for CP Cru from expansion parallax: Downes \& Duerbeck
2000).   Relative to Paper I (see previous footnote) there are a
few changes in Table 1 due to our improved knowledge
of the times of visual maxima, thanks to diligent searches of the
records of Southern hemisphere amateurs (F. Bateson \& A. Jones,
private communication).

\section{Spacecraft and Instrument}

\subsection{Introduction to Gamma-Ray Analysis}

In space-based $\gamma$-ray astronomy the signal from any cosmic source 
is usually completely dominated by intense ambient background 
radiation arising from bombardment by the all-pervasive energetic cosmic 
ray particles, and by the ensuing secondary radiations such as
$\beta$-decays.  The central problem that all experiments must face is
therefore how to subtract away this huge background (typically
$>99\%$ of count rate).\footnote{ It is particularly unfortunate
for this analysis that Be is a common choice for space structures
(\S \S 3.2, 4.1), being a relatively small source of background at energies 
{\em other than\/} 478 keV).}  

In general, the signal has to be modulated.  Methods
include pointing on and off source rapidly (as with 
CGRO/{\em OSSE\/}, Johnson et al. 1993), modulation in time by a 
periodic occulter (e.g. an opaque aperture stop, or the spinning 
spacecraft body like {\em HEAO-3\/}: Mahoney et al. 1982), or even Earth 
(CGRO/{\em BATSE\/}: Ling et al. 2000), or modulation in space by a coded
mask illuminating a segmented detector plane (to be used by 
{\em INTEGRAL\/}/SPI: Vedrenne et al. 1999).  It is important 
to note that our work with TGRS used two
distinct methods of background subtraction.  One is a mechanical 
occulter (\S 5), but the other is unique to TGRS --- unlike any
other experiment TGRS can model its background lines and their
temporal behavior accurately enough to perform a usable background
subtraction (\S 4).

The two methods of background subtraction yield
results for different "samples" of novae.  Occultation measures the
averaged collective $^{7}$Be emission from the general direction of the
GC.  Background modeling ought to detect sudden increases in an
otherwise predictable background line time history, each the 
signature of an individual nova.  We can search the whole time history,
not just the times of the known novae in Table 1, in the hope of
detecting nearby undiscovered events.

\subsection{Instrument and Detector}

The TGRS experiment is located on the south-facing surface
of the rotating cylindrical {\em Wind\/} body, which points permanently
toward the southern ecliptic pole.  The spectrometer, a radiatively cooled
35 cm$^{2}$ n-type Ge crystal sensitive to energies between 20 keV--8 
MeV, is kept at its operating temperature 85 K by a passive radiative
cooler constructed mainly of Be and Mg.  Apart from 30 mm of Be/Cu 
alloy that absorbs solar X-rays, it is unshielded, so that in normal 
(non-burst) operation background
spectra are accumulated continuously from the entire southern ecliptic 
hemisphere.  A 1 cm thick Pb occulter attached to the {\em Wind\/} body
exploits the 3 s rotational period in order to modulate the signal from 
part of the sky.  The occulter subtends $90^{\circ}$ along the ecliptic 
plane,\footnote{ The zenith angle of the occulter mid-plane relative to
the detector is in fact $84.5^{\circ}$ rather than exactly the plane of
the ecliptic, which would be $90^{\circ}$.  The $5.5^{\circ}$ offset
was designed to optimize modulation of the GC.} as seen 
from the detector, the width of the band thereby occulted being 
$16^{\circ}$ FWHM at around 500 keV.

Owens et al. (1995) reported the results of extensive Monte Carlo
simulations of the instrument response as a function of energy and
zenith angle.  The actual cross section 35 cm$^{2}$ is customarily
reduced to an "effective" area by factoring in the efficiency of
detection at a given energy and angle.  Further, the frequent cases
where photon energy is only partly lost in the detector cannot be
used in the present analysis, since the counts appear in a smooth
quasi-continuum at energies below the line energy, and no one count
can be unambiguously associated with a weak line like that at 478 keV.
Our analysis therefore only uses photons which undergo full energy
loss (where in any case 
the Ge response as a function of energy loss is maximal), i.e. which 
appear in the channel(s) at the nominal line energy and are
referred to as the "photopeak".  In these terms, the photopeak 
effective area of the TGRS detector is $\simeq 15$ cm$^{2}$ at
478 keV and zenith angle $0^{\circ}$.  It falls sharply with increasing
energy but does not vary much with zenith angle (Owens et al. 1995).
The instrument response was operationally verified by on-orbit
comparisons of $\gamma$-ray burst spectra with those obtained by
CGRO/{\em BATSE\/} (Seifert et al. 1998) and {\em Wind\/}/KONUS 
(Palmer et al. 1996).

The sharp energy resolution of the detector is critical for our analysis.
Scintillator detectors like {\em SMM\/}/GRS and {\em BATSE\/} typically
achieve $\sim 40$ keV resolution around 500 keV.  Any line at 478 keV is
therefore blended with the positron annihilation line at 511 keV, which
is much stronger in both background and background-subtracted spectra.
The resolution of TGRS at this energy was nominally about 3 keV FWHM,
which was achieved in the early months of the mission (Harris et al.
1998); thereafter resolution degraded due to accumulated damage
from cosmic ray impacts.  The line profiles also became distorted, with
marked tails on the low-energy wings.  
We limited the data analysis to the period
before fall 1997, after which the deterioration became marked (Kurczynski
et al. 1999).  The detector gain stability also deteriorated on this
time-scale.  Although the gain shift was not difficult to monitor using
the positions of known narrow background lines, the effect on the 
analysis of occulted data was serious (\S 5.1).

The TGRS background spectra were binned in 1 keV energy channels over
the entire 20 keV -- 8 MeV range in 24 minute intervals.  The occulted
spectra were accumulated in 128 angle bins covering each 3 s revolution
of {\em Wind\/}.  Telemetry constraints did not allow all of these 
occulted data to be returned in finely-resolved energy bins, however.
Data above 1~MeV were discarded.  
Four energy windows were defined to be 20--150 keV, 150--400 keV, 
400--1000 keV, and 479--543 keV, with energy binnings of 2, 4, 8 and 1
keV respectively.  The boundaries of the windows changed during the
mission as a result of the gain shift.

The unusual orbits followed by {\em Wind\/} since launch provide the
key to the success of the TGRS background modeling.  They have
all been highly elliptical (Acu\~{n}a et al. 1995), and TGRS has 
thus spent
virtually the whole mission in interplanetary space, avoiding major
sources of $\gamma$-ray background radiation such as Earth albedo
$\gamma$-rays and energetic trapped charged particles.  The chief
source of particle bombardment is the Galactic cosmic ray (GCR) flux.
There have been very few interruptions in the data stream, 
generally caused by brief passages of the trapped radiation 
belts around perigee and by triggering of a special mode of data
collection by solar flares and $\gamma$-ray bursts.  The 
live time in this data set covers about 90\% of the analysis period
(Paper II).

The resulting TGRS background lines may be strong, but they are stable 
enough for our purposes (\S 4 below).
By contrast, experiments in low Earth orbit are plagued by 
trapped radiation belt-induced radioactivity, generating background
lines varying on a wide variety of $\beta$-decay time-scales, many
of which never reach equilibrium, and cannot be predicted on either
long or short time-scales (Weidenspointner et al. 2001a).

\section{Analysis of Background Data}

\subsection{Qualitative Features}

Our analyses of the TGRS background and occulted data sets used quite
distinct methodologies that were designed to measure different 
quantities.  Here we focus on the background spectra.  A background 
spectrum is composed of many narrow lines from radioactivity induced 
by GCR and secondary neutron impacts, superimposed on a
continuum.  The large mass of Be close to the TGRS detector in the
radiative cooler and sunshield (\S 3) gives rise to a strong line at exactly
478 keV from the interactions $^{9}$Be($p$,$pnn$)$^{7}$Be and
$^{9}$Be($n$,$3n$)$^{7}$Be (Naya et al.
1996).  We measured the strength of this line in successive spectra 
from the years 1995--1997.  Any sudden increment in the strength would
correspond to the signal from a cosmic source of $^{7}$Be, and would
be confirmed by a subsequent decline on a 53 d half-life.  This
analysis cannot detect a DC level of 478 keV line emission.

This method clearly relies on the underlying background 478 keV line
being very stable, or at least varying in a simple and comprehensible
way.  As mentioned above, the predominant factor has been radioactivity
induced by the GCR flux which varies slowly with the
11 yr solar activity cycle.  Solar magnetic modulation is
expected to cause the flux to peak around solar activity minimum, which
occurred about halfway through our 1995--1997 analysis period.
We found evidence for this in the count rate for partial energy
losses $>~8$ MeV in the detector that is shown in Fig. 1 (top).
This is above the energy region characteristic of $\gamma$-rays produced
by nuclear interactions, so that the background must be due principally 
to GCR bombardment and the diffuse cosmic background, of which the 
former only is time-variable.  The day-to-day variability  of the $>~8$
MeV rate (top full line) shows a very weak $\sim 14$
d period that is caused by the {\em Wind\/} orbit crossing the
neutral current sheet in the interplanetary magnetic field (which
varies with the 28 d solar rotation).  When integrated over the 53 d
time-scale characteristic of nova signals, the count rate shows a 
modest, smooth variation (histogram) that follows the trends expected for 
the GCR flux when subject to solar modulation.  This
pattern of variability was confirmed by a large number of measurements of
TGRS background line strengths performed for line identification 
purposes (Weidenspointner et al. 2001b).  Many
prompt de-excitation lines were identified that arise from 
electromagnetic decays of excited nuclear states on submicrosecond
time-scales, and it is clear that the intensities of these lines
must trace the incident GCR flux effectively instantaneously.  These
trends are also shown in Fig. 1 (full lines).

\subsection{Analysis}

The background spectra were summed in 53 d intervals.  In each interval
the energy range 460--490 keV was fitted with a spectral model consisting
of a power law and two lines.  There is a strong line at 472.2 keV due
to $^{24m}$Na from spallation of structural Al, and the 477.6 keV
$^{7}$Be line from the radiative cooler.  The $^{24m}$Na line was
slightly distorted by the low-energy tailing problem even before the fall
of 1997 (\S 3.2).  It was therefore fitted by an asymmetric function
\begin{eqnarray}
A_{0}~\exp{(-\frac{E-E_{0}}{\sigma ^{2}})}~+~A_{1}~\exp{(
\frac{E-E_{0}}{\nu})}~erfc(\frac{E-E_{0}}{\sigma}~+~\frac{
\sigma}{2 \nu})
\end{eqnarray}
for amplitudes $A_{0}$, $A_{1}$, line energy $E_{0}$ and width 
parameters $\sigma$
and $\nu$ (Phillips \& Marlow 1976).  The line at 478 keV was not
fitted by this function, since other factors broaden it (symmetrically) 
to such an extent that the asymmetry is not significant.  One such
factor is intrinsic;  this line is measured to be somewhat 
broader than the instrument resolution due to blending with a weak line
of $^{55}$Co.  Alternatively, folding its profile with a 
Gaussian corresponding to 1000 km s$^{-1}$ 
(typical for CO novae: \S 2) has the same effect.
A simple Gaussian with this width was therefore included in the fit.  
A specimen of such a fit is shown in Fig. 2.

The amplitude in terms of count rate was corrected for efficiency 
using the photopeak effective area as a function of energy and
zenith angle (\S 3.2).  The zenith angles of the five known 1995--1997
novae were readily obtained; in the general case
(searching for a flux increase when no nova was known to be present) we
assumed a zenith angle of $60^{\circ}$, which
is the average for the estimated distribution of southern novae
(Hatano et al. 1997).

\subsection{Results}

The time series of the line amplitudes fitted to the 478 keV line 
in the background spectra is shown in Fig. 3 ({\em top\/}).  We fitted 
this time series with a model by which $^{7}$Be is created
in the instrument by a time-dependent GCR flux, and decays with a
half-life 53.28 d.  The various curves in Figure 1 suggest that a 
simple parametrization
of the GCR flux would be a linear increase between 1995 January and 
solar minimum in 1996 June,\footnote{ The epoch of solar minimum
differs slightly between the various measured quantities (including
our own) by which it is defined.  Sunspot minimum corresponds to 1996 
June.  Other possible epochs lie between 1996 May--August.} 
followed by an abrupt change of slope.  Let the corresponding $^{7}$Be
production rate at time $t$ be $R~+~Ct$ where $R$ and $C$ are constants
and $C$ changes slope at solar minimum.  The time-dependent $^{7}$Be
abundance is then
\begin{eqnarray}
N(t)~=~(\frac{R}{\lambda}-\frac{C}{\lambda^{2}})~[1-\exp{(-\lambda t)}]~
+~\frac{C t}{\lambda}~+~N(t=0)~\exp{(-\lambda t)}
\end{eqnarray}
where $\lambda = 76.87$ d.  We fitted the time series
to this function (which has 5 free parameters when account is taken
of the discontinuity at solar minimum) and searched for significant
sudden flux increases over and above it.  This was done by fitting
an exponentially decaying function with decay constant $\lambda$
to each residual data point and the following ones.  The fit using
Eq. (5) is shown in Fig. 3 ({\em top\/}), and the residuals
in Fig. 3 ({\em bottom\/}).

It is evident from  Fig. 3 ({\em top\/} and {\em bottom\/}) that
the function (2) fitted the data very well, so the procedure of 
fitting the exponential decay could be applied without obvious systematic
errors.  The clear result in Fig. 3 ({\em bottom\/}) is that no flux 
increase followed by exponential decline
was found to appear in the 478 keV line, either at the times of
known southern hemisphere novae or at any other time.
Clearly no hitherto undiscovered novae were detected by $\gamma$-ray
line emission using this method.  The $3 \sigma$ upper flux limits
for this general case, and for the five known novae, are given in
Table 2.

\section{Analysis of Occulted Data}

\subsection{Analysis}

Our analysis of the occulted data was designed to measure a 
quasi-steady, quasi-diffuse
478 keV $\gamma$-ray line arising from the integrated emission of many
novae that would have occurred towards the GC during the 
53 d half-life of $^{7}$Be.  The principle was suggested by Leising
(1988), and was put into practice by Harris, Leising \& Share (1991).

Our procedure was essentially identical to that which we previously
employed to measure the diffuse positron annihilation spectrum from
the GC (Harris et al. 1998).  When the count rate in
an energy channel is
plotted as a function of angle along the ecliptic, those angle bins
corresponding to the direction of a source show dips.  The amplitude
of the dip equals the source count rate at that energy, so that a
spectrum of the source is built by taking the source count rate dips 
in successive energy channels.  It is necessary to assume a model of
the source's spatial distribution; we chose a Gaussian in ecliptic
longitude of FWHM $24^{\circ}$, which Harris et al. (1998) found to
be the best estimate for the distribution of the diffuse 511 keV 
annihilation line (which is compatible with a nova-like distribution of
sources: Milne et al. 2001).

The energy channel binning in the region of interest to us
around 478 keV was non-uniform and time-variable.  This is because,
as noted in \S 3.2, data were only returned at 
the optimum 1 keV binning in a single 64-keV wide energy window 
centered approximately on 511 keV, and unfortunately the 478 keV line 
was on the edge of this window.  Energy resolution in the occulted data was 
therefore bad (8 keV) on the red side, and moreover the transition between
well- and coarse-binned data varied during the mission as the 
high-resolution window shifted with the instrument gain. 

Rather than using the complete 1995--1997 Harris et al. (1998) TGRS 
spectrum, we used a series of spectra summed over 90 d intervals
which Harris et al. also generated (for their search for 511 keV line
variability).  This time-scale happens to be similar to that for
significant variability of the gain.  We could thus treat the
gain shift as a constant and "fix" the location of the high-resolution 
window boundary in a manner that facilitated spectrum modeling. 
The 90 d spectra in the range 460--490 keV 
were fitted by a model consisting of a power law plus a Gaussian line 
at 478 keV.  A typical example of this fit is shown in Fig. 4.  Note 
the abrupt transition from coarse to fine binning in the middle of the 
energy range.

\subsection{Results}

The measured 478 keV line fluxes during 90 d intervals are shown in
Fig. 5.  These results may be combined to give the total measured
quasi-steady flux from the GC.  It is clear that no line
is detected, within limits which are given in Table 2.  The 
sensitivity of this search is degraded by a factor 4--5 relative
to that of the background-modeling method.  This is mainly due to
the gain shift problem mentioned above.  Table 2 also shows
earlier measurements by {\em SMM\/} for both individual novae and
the quasi-steady integrated GC flux (Harris et al. 1991).
Note that the integrated flux measurements are not exactly
comparable, since the {\em SMM\/} value comes from a very broad
region $\sim 130^{\circ}$ across around the GC, whereas the
TGRS value reflects the signal from the $16^{\circ} \times 90^{\circ}$
occulted region only.

\section{DISCUSSION}

\subsection{Comparison of TGRS Results with Theory}

From the flux upper limits in Table 1 we calculate the $^{7}$Be
abundances in the sources, given the distances.  
Our limits improve upon the {\em SMM\/} values 
by an order of magnitude for individual events, and by a factor 2 for 
the integrated flux from the GC.  Unfortunately we do not obtain proportionate
improvements in the limits on the $^{7}$Be abundances from these
fluxes, for two reasons.  In the case of individual novae, by misfortune 
the TGRS sample lay at greater distances than the {\em SMM\/} sample.  
In the case of the GC, the measurement methods
({\em SMM\/} or TGRS occulter) return $^{7}$Be abundances integrated
over the unknown number of novae that are simultaneously present, and
thus depend on the uncertain Galactic nova rate $R_{N}$ and on the 
aperture of the instrument.
Since the effective aperture of TGRS is smaller than the $\sim 130^{\circ}$
aperture of {\em SMM\/}, the upper limit implied by a given $\gamma$-ray
flux is divided among fewer novae.   

Comparisons with theory must take into account the shortcomings and
uncertainties in nova models (see \S 2).  For our purposes, the 
prediction of any nova's $^{7}$Be mass may be factored into the product
of the ejected mass $M_{ej}$, and the mass fraction $X_{7}$ of $^{7}$Be
in it.  In the favorable case of a massive CO nova, "optimistic" models such 
as G\'{o}mez-Gomar et al. (1998) suggest values $M_{ej}/M_{\sun} \sim X_{7}
\sim 10^{-5}$.  Both of these values may be underestimates, leading to
yet more optimistic scenarios that were proposed earlier by
Clayton (1981).

There are two arguments for this kind of scenario.\footnote{ We assume 
that the factors $M_{ej}$ and $X_{7}$ can be varied independently.  This is 
unlikely to be the case in actual models, although Starrfield et al. (1997)
made a similar assumption in order to generate $\gamma$-ray predictions.}  
First, as
pointed out in \S 2, there is a "missing mass" problem (Starrfield et
al. 2000b): nova models in general predict $M_{ej}$ that are much too
low.  Measured ejecta masses range from $\sim 10^{-5}$--$10^{-3}~M_{\sun}$
(Warner 1995); the mean value is probably $\sim 2 \times 10^{-4}~M_{\sun}$
(Gehrz et al. 1998).  As for $X_{7}$, it was pointed out by Starrfield
et al. (1978) that, whereas models assume that the abundance of $^{3}$He
in the gas accreted onto the white dwarf is usually solar, this is
probably not the case.  The $^{3}$He abundance should be enhanced in the 
donor star due to synthesis of this isotope in the pp chain,
and convection will mix the $^{3}$He-rich material into
the surface layers which are transported to the white dwarf.
Older main sequence star models suggested $^{3}$He enhancements 
$\sim 10$ times solar, while more recent models suggest a value
about 50\% of this (Morel et al. 1999).  However, Boffin et al. 
(1993) showed that, contrary to what was assumed by Starrfield et 
al. (1978), the yield of $^{7}$Be from $^{3}$He($\alpha$,$\gamma$)$^{7}$Be
in novae is not proportional to the
$^{3}$He abundance, but to its logarithm.  The model yields of
$^{7}$Be are then probably enhanced by less than a factor 2.5 by
the $^{3}$He enhancement.  Most of the possible variation in the
product is thus due to uncertainty in $M_{ej}$, of which our upper 
limits may be a test.

For the purpose of comparison with our upper limits, we show in
Table 2 the fluxes that would be expected from each nova (and from
the GC) if optimistic values $M_{ej} = 2 \times 10^{-4} M_{\sun}$
and $X_{7} = 2.5 \times 10^{-5}$, similar to the old values of
Clayton (1981), are assumed.  In all cases, our limits lie above
the expected fluxes; however some of the individual cases are 
of interest.  Our limit for BY Cir is only a factor $\sim 6$ above
the expected value, even better than the {\em SMM\/} limit for
the much closer event V482 Cen.  These happen to be the only
verified CO novae in the two samples.  Thus we can exclude highly
optimistic estimates of the ejecta masses, exceeding 
1--2$\times 10^{-3}~M_{\sun}$, for two members of 
this class of nova.  Given
that CP Cru is the only ONe event in the TGRS sample, with a limit
on the flux 40 times the expected value, the same
reasoning produces an upper limit $M_{ej} < 8 \times 10^{-3}~M_{\sun}$
for this class of event, improving on the {\em SMM\/} limit by almost
a factor 10.  From our result for the general case,
we also exclude any undiscovered nova with optimistic 
"Clayton-like" parameters having occurred within 1.1 kpc
during 1995--1997; or any undiscovered nova of the highest-yielding
G\'{o}mez-Gomar et al. (1998) type (CO, $1.15~M_{\sun}$) within 134
pc.

\subsection{Possible Application to {\em INTEGRAL\/}}

This work has established the feasibility of the time series modeling 
method of background subtraction (\S 4), at least for resolvable lines 
(not too severely blended),
for $\gamma$-ray emitters having appropriate time-scales --- and 
especially for stable background lines amenable to very simple 
temporal models.  An obvious improvement would be to model blended
line complexes of astronomical interest, which would require much
more sophisticated modeling than our semi-empirical approach.  Thus,
work is under way to model the TGRS background spectrum ab
initio, from the mass model and the GCR flux
(Weidenspointner 2001c).

Some of the features that make this approach possible are included in
the upcoming {\em INTEGRAL\/} mission, in particular a high-resolution
Ge spectrometer (SPI), and a highly elliptical orbit which, like that
of TGRS, avoids Earth's trapped radiation belts (Winkler 1996).  If
the background lines prove to be stable enough, it may be possible to
apply our method to the off-source SPI background.

Relative to TGRS, SPI has much greater sensitivity, close to $10^{-5}$
photon cm$^{-2}$ s$^{-1}$ at 478 keV.  The best-case theoretical models 
in Table 2
(CO, $1.15~M_{\sun}$) are then detectable at a distance of 500 pc
(Hernanz \& Jos\'{e} 2000).  The more optimistic parameters 
$M_{ej} = 2 \times 10^{-4}~M_{\sun}$, $X_{7} = 2.5 \times 10^{-5}$,
similar to those of Clayton (1981), that we discussed in \S 6.1 and
assumed in column 5 of
Table 2 for the purpose of flux comparisons, would enable SPI to
make detections out to 3.4 kpc.  A detection distance of this order
intercepts $\sim 0.8$\% of the global nova distribution (Paper II),
two-thirds of which are of CO subtype.  If {\em INTEGRAL\/} continues
beyond its 2-year nominal mission to the 5-year extended mission, the
number of novae it can be expected to detect is $\sim 0.03 R_{N}$,
which is $\sim 1.3$ for realistic values of $R_{N} \sim 50$ yr$^{-1}$.
The prospects for a SPI observation improving on our results are
therefore quite good.

\acknowledgements

We thank Albert Jones and Frank Bateson of the Royal New Zealand 
Astronomical Society for prediscovery limits on some of the 
novae in Table 1.  We are grateful to Theresa Sheets (LHEA) and
Sandhia Bansal (HSTX) for their work on the analysis software.

\clearpage

\begin{figure}

\caption{Variation of 
monitors of GCR intensity through 1996 solar minimum.
(Top, full line) Count rate at energies $>8$ MeV on 1 d time-scale.
(Top, histogram) Count rate at energies $>8$ MeV on 53 d time-scale.
(Lower full lines) Intensities of prompt instrumental de-excitation lines as
measured by local fits with power law and Gaussian profile.
Line ID --- 140 keV, $^{75}$Ge(139.7--g.s.) --- 198 keV, 
$^{71}$Ge cascade (198--g.s) sum peak --- 440 keV, $^{23}$Na(439.9--g.s.).}

\caption{TGRS 
background spectrum during 1995 July 31--1995 September
22, fitted by power law and lines at 472 keV ($^{24m}$Na) and 478 keV
($^{7}$Li).}
 
\caption{({\em Top\/}) Time series 
of 478 keV line strengths in 53 d spectra, fitted to a model of
the behavior of the instrumental background line (full line).  Arrows --- 
estimated eruption times of known southern novae: in sequence, BY Cir,
V888 Cen, V4361 Sgr, CP Cru, and N Sco 1997. ({\em Bottom\/}) Residual
flux after subtracting the above model.  The value of $\chi^2$ per
degree of freedom for the fit is 0.4.}

\caption{GC spectrum 
around 478 keV from TGRS occultation analysis, for 
the period 1995 December 25--1996 March 22, fitted by a hypothetical 
cosmic line at 478 keV superimposed on a power law.  Line 
amplitude is $-4.7 \pm 7.5 \times 10^{-5}~\gamma$ cm$^{-2}$ s$^{-1}$.}

\caption{Time series of GC fits to 90 d spectra as in Fig. 4.  
The $\chi^{2}$ per degree of freedom for the fit is 1.0.}

\end{figure}

\clearpage

\begin{table*}

\small
\begin{center}
\begin{tabular}{lcccccc}
\tableline
Nova & White & Possible date & $m_{V}$ at & $t_{2}$, & Distance, & Distance, \\
 & Dwarf & of maximum & discovery & days & Eq. (1--3), pc$^{a}$
& $L_{Edd}$, pc$^{b}$ \\
\tableline
BY Cir & CO & 1995 Jan 21.0 -- Jan 27.3 & 7.2 & $\sim 20$ & 3240--4000 & 
2370--3030 \\
V888 Cen & & 1995 Feb 22.0 -- Feb 23.3 & 7.2 & $\sim 6$ & 6060--6970 & 
2450--3720 \\
V4361 Sgr & & 1996 Jun 19.6 -- Jul 11.5 & 10.0 & 53 & 4640--6785 & 
5070--10280 \\
CP Cru & ONe & 1996 Aug 22.0 -- Aug 27.0 & 9.25 & 5.2 & 6770--16300 & 
3100--7500  \\
V1141 Sco & & 1997 Jun 2.1 -- Jun 5.1 & 8.5 & 4.5 & 6830--8340 & 2460--6860 \\
\tableline

\end{tabular}
\end{center}
~  \\
$^{a}$~Uncertainty does not include observational scatter
around the empirical law Eq.(1--3). \\
$^{b}$~Uncertainty does not include bolometric correction at
visual maximum. \\

\caption{Novae observable by TGRS, 1995--1997}

\normalsize
\end{table*}

\clearpage

\begin{table*}
\begin{center}
\begin{tabular}{lcccccc}
\tableline
Target & Distance & Zenith & Flux~$^{b}$ & Expected~$^{c}$ 
& Implied $^{7}$Be mass \\
 & pc$^{a}$ & Angle & $\gamma$ cm$^{-2}$ s$^{-1}$ 
 & $\gamma$ cm$^{-2}$ s$^{-1}$ & $M_{\sun}$ per nova$^{b}$ \\
\tableline
{\em Individual Novae\/} & & & & & \\
Undiscovered nova & & $60^{\circ}$ & $1.0 \times 10^{-4}$ & & \\
BY Cir & 3160 & 45$^{\circ}$ & $6.8 \times 10^{-5}$ & 
$1.1 \times 10^{-5}$ & $3.0 \times 10^{-8}$ \\
V888 Cen & 4800 & $42^{\circ}$ & $6.3 \times 10^{-5}$ &
$4.9 \times 10^{-6}$ & $6.4 \times 10^{-8}$ \\
V4361 Sgr & 6700 & $95^{\circ}$ & $1.1 \times 10^{-4}$ &
$2.5\times 10^{-6}$ & $2.2 \times 10^{-7}$ \\
CP Cru & 3180~$^{d}$ & $37^{\circ}$ & $8.8 \times 10^{-5}$ & 
$2.2 \times 10^{-6}$ & $3.9 \times 10^{-8}$ \\
Nova Sco & 6120 & $97^{\circ}$ & $1.6 \times 10^{-4}$ & 
$3.0 \times 10^{-6}$ & $2.7 \times 10^{-7}$ \\
V1370 Aql~$^{e}$ & 3500 & & $1.2 \times 10^{-3}$ & $1.8 \times 10^{-6}$ &
$6.3 \times 10^{-7}$ \\
QU Vul~$^{e}$ & 3000 & & $7.5 \times 10^{-4}$ & $2.5 \times 10^{-6}$ &
$3.1 \times 10^{-7}$ \\
V842 Cen~$^{e}$ & 1100 & & $9.6 \times 10^{-4}$ & $9.3 \times 10^{-5}$ &
$5.2 \times 10^{-8}$ \\
 \\
{\em GC Integrated\/} & & & & & \\
TGRS & 8000 & $84.5^{\circ}$ & $7.7 \times 10^{-5}$ & 
$7.8 R_{N} \times 10^{-8}$ & $3.4 \times 10^{-6}/R_{N}$~ $^{f}$ \\
{\em SMM\/} & 8000 & & $1.5 \times 10^{-4}$ & $1.6 R_{N} \times 10^{-7}$ &
$3.5 \times 10^{-6}/R_{N}$~$^{f}$ \\
 \\
{\em Theory\/}~$^{g}$ & & & & & \\
CO $0.8~M_{\sun}$ & 1000 & & $1.8 \times 10^{-6}$ & & $8.0 \times 10^{-11}$ \\
CO $1.15~M_{\sun}$ & 1000 & & $2.5 \times 10^{-6}$ & & $1.1 \times 10^{-10}$ \\
ONe $1.15~M_{\sun}$ & 1000 & & $3.6 \times 10^{-7}$ & & $1.6 \times 10^{-11}$ 
\\
ONe $1.25~M_{\sun}$ & 1000 & & $2.7 \times 10^{-7}$ & & $1.2 \times 10^{-11}$ 
\\

\tableline

\end{tabular}
\end{center}
~  \\
$^{a}$~Simple mean of estimates in columns 6 and 7 of Table 1, unless
otherwise noted. \\
$^{b}$~$3~\sigma$ upper limit. \\
$^{c}$~Assumes "optimistic" parameters (see \S 6) of $2 \times 10^{-4}~M_{\sun}$
of ejecta (as observed: Warner 1995) enriched to mass fractions 
$2.5 \times10^{-5}$
for CO subclass and $5 \times 10^{-6}$ for ONe subclass. \\
$^{d}$~From expansion parallax measurement (Downes \& Duerbeck 2000). \\
$^{e}$~Measured by {\em SMM\/} (Harris et al. 1991).  
Subclasses: ONe (V1370 Aql 1982 and QU Vul 1984), CO (V842 Cen 1986).  \\
$^{f}$~The effective apertures of TGRS and {\em SMM\/} will 
contain respectively $0.06~R_{N}$ and $0.12~R_{N}$ novae at any one time, 
where $R_{N}$ is the total Galactic nova rate per year.  The total 
$^{7}$Be masses inferred from the flux upper limits
have been divided by these factors to get the yield per nova. \\ 
$^{g}$ Hernanz \& Jos\'{e} (2000) and G\'{o}mez-Gomar et al. (1998).
In column 2 1 kpc is an arbitrary choice for the purpose of line flux 
comparisons.  In column 4 the ejecta masses are taken to be the actual
model values 1--6~$\times 10^{-5}~M_{\sun}$, not the "optimistic" value 
$2 \times 10^{-4}~M_{\sun}$ assumed elsewhere (see note {\em c\/}).\\
 \\
\caption{Results for 478 keV line fluxes and $^{7}$Be yields.}

\end{table*}

\clearpage

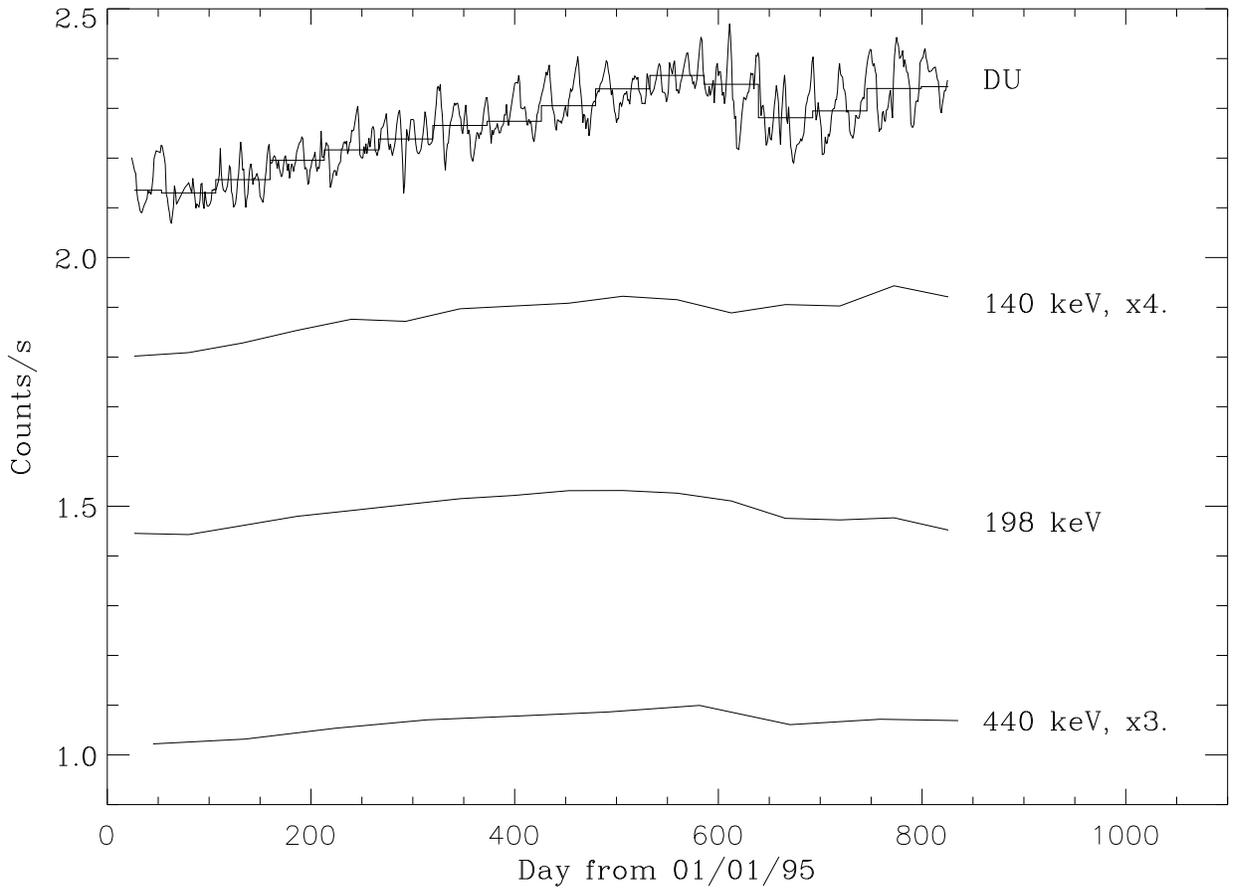
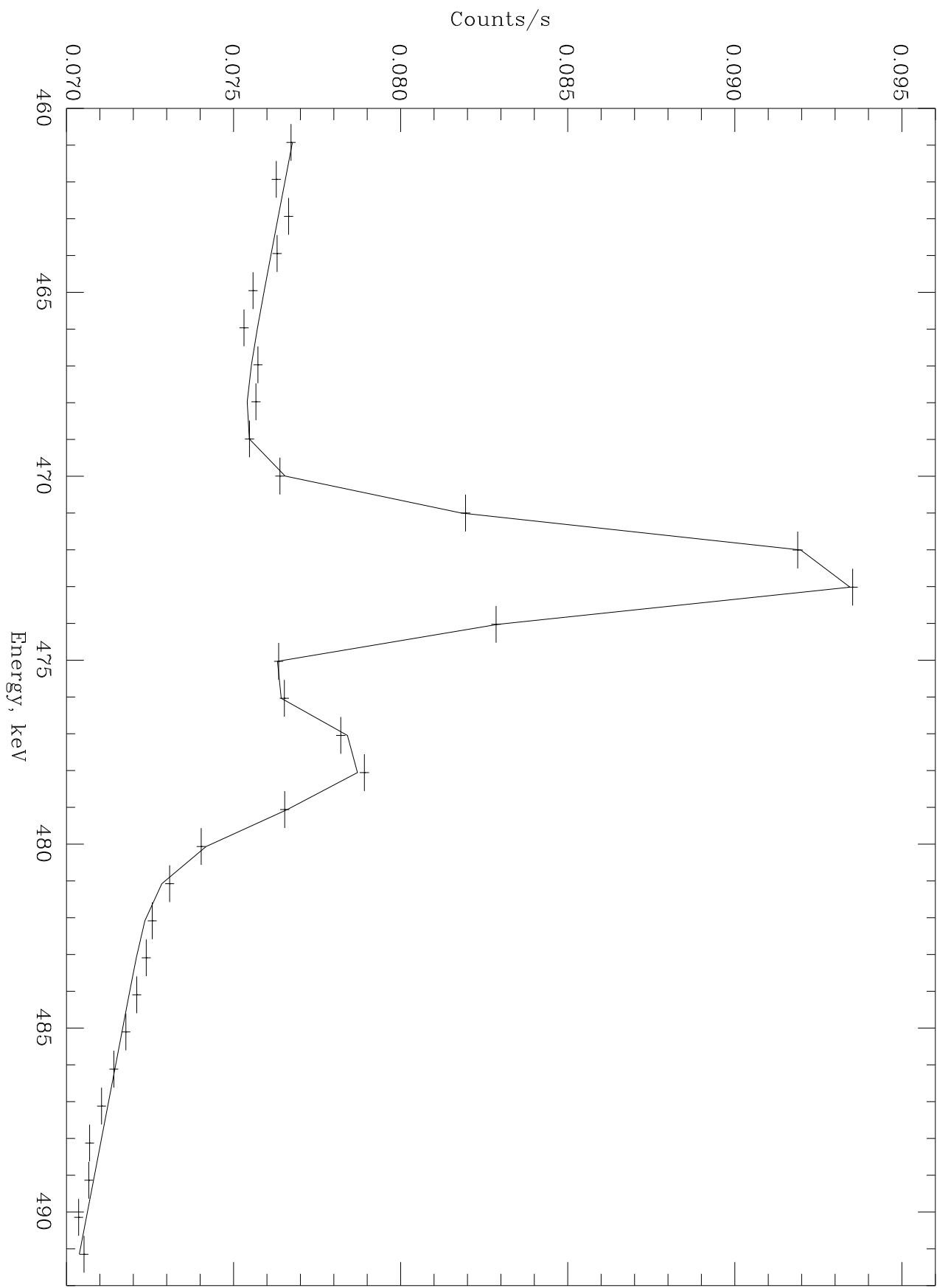
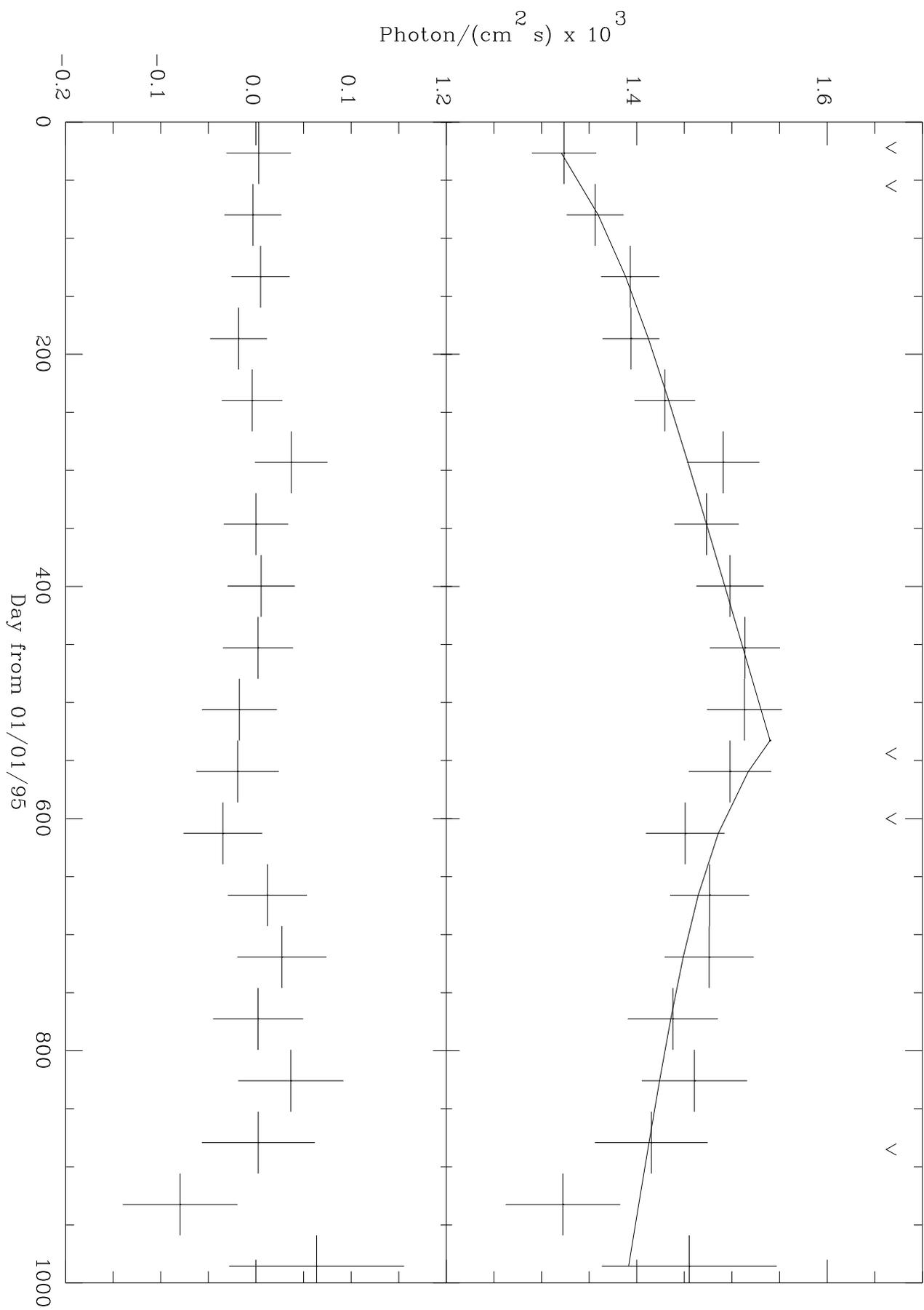
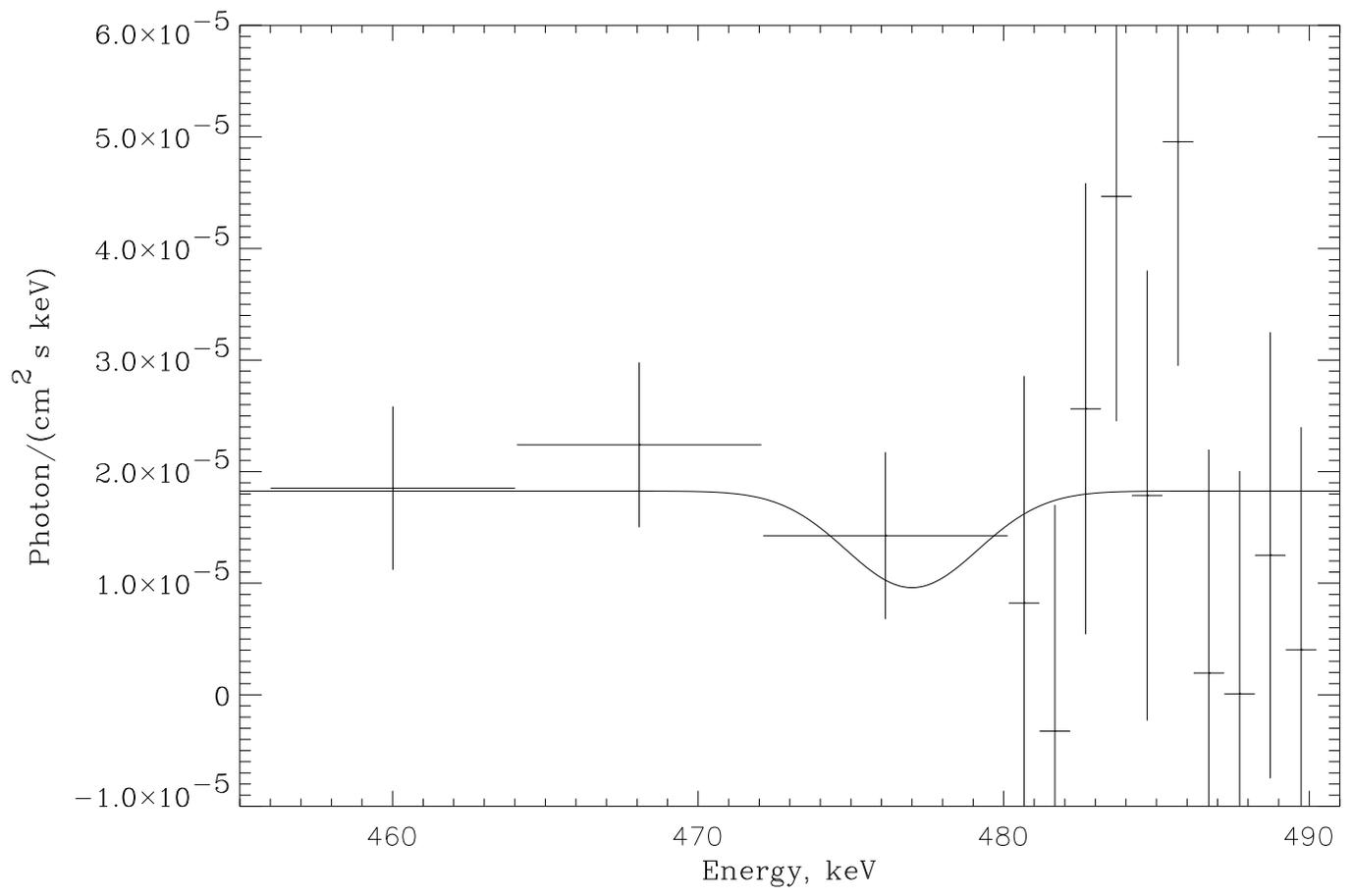
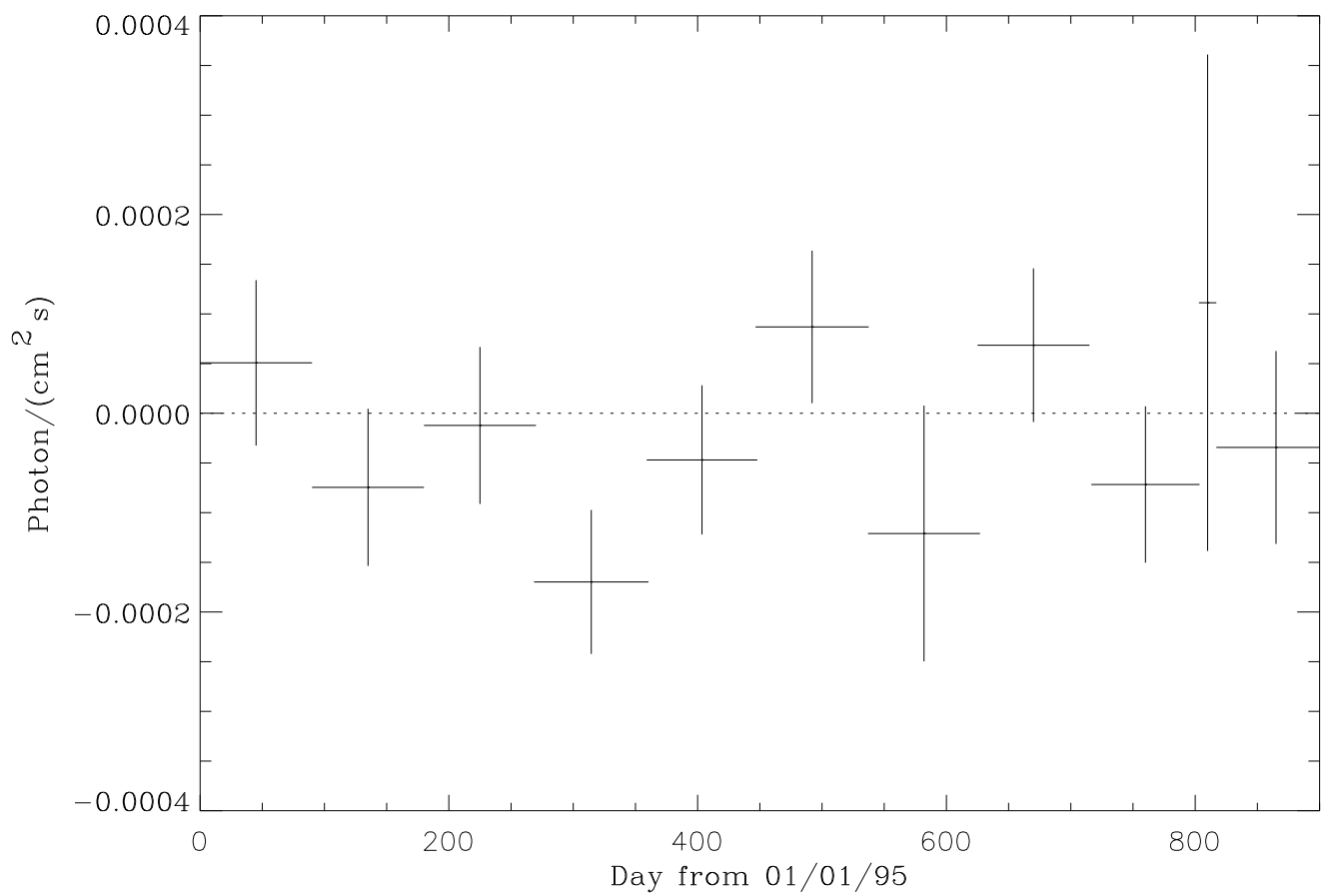
\end{document}